# Propuesta para la mejora en la transferencia y ejecución de múltiples instancias de una imagen virtual

**Tomás Ramírez Picarzo**[*]
Universidad de Extremadura, España

**Francisco Fernández de Vega**[**]
Centro Universitario de Mérida. Universidad de Extremadura, España

**Daniel Lombraña González**[***]
Citizen Cyberscience Centre (CCC), Switzerland

**Resumen.** La tecnología de virtualización permite en la actualidad ejecutar cualquier aplicación compleja y de alto coste computacional (las aplicaciones científicas son un buen ejemplo) sobre sistemas distribuidos heterogéneos, que hacen uso habitual de tecnologías Grid y Cloud, permitiendo así un ahorro significativo de tiempos de cálculo. Este modelo es particularmente interesante para la ejecución masiva de simulaciones y cálculos científicos, ya que permiten la ejecución paralela de aplicaciones utilizando el mismo entorno de ejecución (sin modificaciones) que utiliza el científico de modo habitual. Sin embargo el uso y distribución de imágenes virtuales de gran tamaño puede ser un problema (hasta decenas de GBytes), que se agrava cuando se pretende un envío masivo sobre un número alto de computadores distribuidos. Este trabajo tiene como objetivo principal presentar un análisis del modo de ejecución y una propuesta para la mejora (reducción en el tamaño) de las imágenes virtuales pretendiendo aminorar el tiempo de distribución en sistemas distribuidos. Para ello se realiza un análisis muy específico de los requerimientos que necesita un sistema operativo (SO invitado) en algunos puntos de su ejecución.

**Palabras clave:** virtualbox, virtualización, Grid, Cloud, Computación Voluntaria.

## 1. Introducción

Existen en la actualidad proyectos a nivel internacional (cientos de ellos de carácter científico: Einstein@home[1], ExtremaduraAtHome[2], etc.) donde miles de máquinas trabajan para una misma tarea. La diversidad de los sistemas heterogéneos conectados tiende a ser un obstáculo para los desarrolladores de software, que pretenden abarcar la amplia diversidad de sistemas o entornos computacionales. Igualmente esta heterogeneidad podría suponer en muchos casos un problema para sincronizar o unificar la capacidad de cómputo de las redes interconectadas ya que en la actualidad las comunidades científicas tienen la necesidad de intercambiarse constantemente sus diversos proyectos con la intención de compartir y trabajar conjuntamente en las mismas tareas.

La tecnología de virtualización permite disponer en una imagen virtual del entorno completo de ejecución de cualquier proyecto: SO (sistema operativo) y aplicaciones. En estas imágenes virtuales se incluyen proyectos, aplicaciones científicas, configuración del entorno de ejecución, etc. de manera que aseguren completamente su funcionalidad. La integración de la tecnología de virtualización en los sistemas distribuidos (Grid, Cloud, etc.) ha permitido, entre otras cuestiones, reducir el impacto de la heterogeneidad de los sistemas,

---

facilitando la administración y despliegue de servicios en diferente hardware. Sin embargo disponer de una máquina virtual (MV) supone en muchos casos la creación de un archivo que por lo general adquiere un gran tamaño (aunque haya otras opciones de creación de MV).

El objetivo principal de este trabajo consiste en analizar la posibilidad de reducir el tamaño de una imagen virtual en base a la utilización de los recursos del SO alojado, es decir, qué partes del sistema de ficheros podrían seleccionarse para obtener exclusivamente lo necesario en la ejecución de un entorno virtualizado específico. Para ello se establecen tres pasos:

- Monitorización del SO alojado.
- Extracción de las partes del SO.
- Generación de instancias.

El resultado de estos tres pasos resulta en una nueva imagen virtual formada exclusivamente por los archivos monitorizados y que por ello mismo, tiene un tamaño bastante inferior al que corresponde al sistema de ficheros completo del SO alojado. De esta forma no se comprime el SO íntegro, como sería el caso de las herramientas genéricas que comprimen (en un alto porcentaje) imágenes virtuales y que en definitiva la compresión que realizan normalmente es a nivel de sectores. Una reducción del tamaño de una MV representa de manera implícita una mejora en la transferencia de las imágenes (latencia). Integrar un procedimiento que permita la reducción del tamaño de las imágenes virtuales en un entorno distribuido, podría mejorar notablemente las transferencias existentes en esa red de millones de ordenadores conectados por todo el mundo y que computan diariamente miles de unidades de trabajo que corresponden a los proyectos en los que colaboran.

La idea de minimizar los recursos de un sistema operativo siempre es un tema de controversia debido a que puede entenderse que no todos los recursos o aplicaciones del propio sistema son necesarios en todas las sesiones que una aplicación científica pudiera necesitar. Los resultados obtenidos en este análisis permiten evaluar algunas propuestas de mejora para la distribución de imágenes virtuales, basándose en la reducción de tamaño del sistema de ficheros incluido en el SO alojado de una máquina virtual.

Por nuestra parte la motivación para realizar este trabajo, de carácter exclusívamente académico, es proponer una línea de estudio en base a que no conocemos otros textos relacionados y que se hayan centrado en la funcionalidad del SO para obtener una reducción de la imagen virtual original e igualmente tengan la intención de aprovechar este beneficio para integrarlos en infraestructuras relacionadas con la computación distribuida, obteniendo así un mayor rendimiento en las tareas de transferencia o distribución de las imágenes.

En la primera sección se introduce al problema que se intenta resolver. El uso de la tecnología de virtualización en la distribución de máquinas virtuales ocupa la segunda sección y pretende presentar con ello ejemplos actuales de entornos distribuidos donde la transferencia de los mismos es fundamental para su optimización. Finalmente las secciones tres y cuatro se dedican a la explicación del análisis propuesto como mecanismo de reducción de imágenes virtuales.

## 1. Distribución de máquinas virtuales.

En esta sección se pretende dar a conocer cuáles son actualmente algunas de las arquitecturas o plataformas que en la tecnología de virtualización, proponen soluciones sobre la distribución de entornos virtuales. El análisis que se realiza en el presente trabajo se encuentra directamente relacionado con estas infraestructuras, ya que cada una de ellas representa una técnica de transferencia de entornos o aplicaciones virtuales donde su tamaño supone un aspecto muy importante. La reducción del volumen de datos transferidos mejora entre otras cuestiones, la distribución del cómputo de trabajo en paralelo entre MV en un Grid o en Cloud y también la adquisición de un entorno virtualizado que suponga, por ejemplo, la ejecución de una aplicación científica.

Por tanto con los siguientes trabajos relacionados en esta sección se pretende repasar algunas técnicas relacionadas con el uso y aplicación de la tecnología de virtualización en entornos diversos como pueden ser: los sistemas distribuidos, las migraciones *live*, etc.

Una tecnología muy conocida y basada en entornos distribuidos o Desktop Grid es la Computación Voluntaria, siendo BOINC una de las herramientas software más conocidas que utilizan este modelo, *David P. Anderson,* [1]. BOINC es un software desarrollado en la Universidad de Berkeley (California) y se inició como un proyecto basado en Computación Voluntaria denominado SETI@home[2]. El objetivo que pretende es que los científicos dispongan de la posibilidad de usar los recursos de ordenadores que forman voluntariamente parte de proyectos relacionados con investigaciones científicas. Tal y como describe el autor *Daniel Lombraña G.* [3]

algunas de las principales características de BOINC son: la autonomía de proyectos, flexibilidad en los voluntarios, marco de aplicación flexible, código abierto, etc. Igualmente otras características de BOINC y que se especifican en *F. Chávez* et al.[4] nos dice que cada proyecto se identifica por una URL maestra (inicio de su sitio web) y donde los usuarios podrán contribuir con tantos equipos como deseen usando el cliente BOINC.

El trabajo de *Segal* et al. [5] hace referencia, entre otras cuestiones a CernVM[3] (*Virtual Software Appliance* para participantes en experimentos LHC[4] del CERN[5]) en relación a una infraestructura de tipo Cloud Computing ("CernVM and Clouds"), así como a sistemas clouds con "PC's voluntarios" controlados por BOINC ("Volunteer Clouds").

Por otro lado existen técnicas de adquisición de entornos virtuales mediante clientes ligeros o "thin-client". Un estudio sobre estas técnicas se puede encontrar en *J. Pujal Curià* et al. [6]. Los clientes ligeros permiten acceder a aplicaciones remotas donde cada usuario se autentica y posteriormente adquiere la imagen que necesite. Otra tecnología basada en el acceso mediante un cliente ligero es la que refleja el trabajo de *R.A. Baratto* et al. [7]. Se presenta *Mobidesk* como una arquitectura mobile virtual desktop y dónde se hace referencia a la movilidad de escritorios de usuario haciendo uso de la tecnología de virtualización.

El estudio ofrecido por *Gabor Kecskemeti* et al. [8] presenta un mecanismo la implementación o instalación de aplicaciones bajo una infraestructura de servicios Grid usando la tecnología de virtualización. Otro trabajo relacionado con la adquisición de aplicaciones en entornos virtuales es el propuesto por *Y. Zhang* et al. [9]. Las aplicaciones se van cargando (por ejemplo desde Internet mediante tecnología P2P, con lo cual se pretende aumentar la velocidad de acceso) en un entorno operativo virtualizado sin necesidad de instalación, teniendo en cuenta que los datos de usuarios personalizados se almacenan en dispositivos portátiles (conservando la privacidad). *C.P. Sapuntzakis* et al. [10] presenta un estudio que tiene como objetivo optimizar las migraciones de un entorno personalizado de trabajo, concretamente, la movilidad de un conjunto de recursos como podrían ser el SO, las aplicaciones, datos, etc. (estado de un entorno personalizado) usando la tecnología de virtualización. Este entorno personalizado se denomina técnicamente "cápsula" (prototipo *"Collective"*). Lo que se pretende es que en una "movilidad" habitual de usuario como puede ser "casa-trabajo-casa" se pueda disponer de un entorno privado de trabajo, concretamente de una cápsula específica en las mejores condiciones técnicas posibles.

En relación a la tecnología de virtualización con migraciones de tipo *live,* destaca el trabajo de *H. Liu* et al. [11]. La tecnología conocida como *CR/TR-Motion* (Checkpointing/Recovery y Trace/Replay) permite de forma transparente y optimizada las migraciones de MV, reduciendo el tiempo de inactividad y el consumo de ancho de banda. Otro mecanismo de migración *live* se explica en el trabajo *R. Bradford* et al. [12] y se centra básicamente en la idea de conservar el estado de las VM una vez que sean migradas sobre redes WAN (wide area network). Para ello propone una implementación (*XenoServer*, *Xen*) basada en una serie de etapas o fases mediante las cuales las MV pueden ser migradas garantizando consistencia, mínima interrupción de ejecución y la transparencia de todo el proceso de migración. Finalmente mencionar las posibilidades que puede aportar el trabajo de *E. Harney* et al. [13], solución propuesta para el estándar *Mobile IPv6* (redes compatibles IPv6) y orientado a las migraciones *live* de VM en redes WAN (Internet).

## 2. Monitorización de los SO alojados.

Una vez que terminado este pequeño repaso a algunas tecnologías relacionadas con entornos virtuales, en esta sección se pretende analizar la propuesta del presente trabajo y para ello se hace necesario comentar previamente sobre algunos aspectos relacionados con el funcionamiento de un SO alojado en una MV.

Los SO (sistemas operativos) que se instalan en los ordenadores o servidores de una organización requieren en cada momento la ejecución de diversas utilidades o servicios que forman parte del mismo (sistema de ficheros). A medida que el usuario del SO va necesitando las diferentes funcionalidades en su sistema (impresión, interfaz de red, ejecución de aplicaciones, etc.) el núcleo o kernel del SO realiza las correspondientes llamadas a estos servicios implementados en el propio SO. Desde hace varios años la ingeniería del software ha evolucionado ampliamente en la teoría y funcionalidad de los SO, tanto a nivel de usuario estándar como a nivel de servidores. Ello ha permitido, entre otras cuestiones, ofrecer versiones reducidas de muchos SO donde la funcionalidad es mínima pero al mismo tiempo suficiente para establecer sesiones de usuario y de servicios implementados en cada SO de una manera exclusiva. Prueba de ello es la diversidad de arquitecturas de SO existentes en la actualidad: versiones *Live* (CD, USB, etc.), SO de tamaño reducido y de funcionalidad limitada (equipos con pocos recursos) o incluso versiones de SO adaptadas a ordenadores portátiles, PDA's, etc. Estos sistemas operativos pueden adaptarse a la funcionalidad que interese al propio usuario del SO y de este modo optimizar el funcionamiento de sus aplicaciones implementadas.

---

3  CernVM. http://cernvm.cern.ch/cernvm/
4  Large Hadron Collider (LHC).
5  CERN (Conseil Européen pour la Recherche Nucléaire). www.cern.ch/

Dado que en el actual trabajo se analiza la posibilidad de reducir un SO en base a sus características de funcionamiento, en esta sección se propone monitorizar la actividad de un sistema operativo alojado en una MV y posteriormente examinar los resultados obtenidos. Se produce por tanto una selección de los datos que se consideran más relevantes y que permite establecer una aproximación al cálculo que supone el tamaño que finalmente podría tener la imagen virtual final.

**3.1 Monitorizar el arranque del sistema.**

El primer punto de ejecución en el SO que se va a monitorizar es el que corresponde al inicio o arranque del sistema. El software utilizado para ello es *readahead*[6]. Para el correcto funcionamiento de este paquete debe configurarse el arranque del SO alojado y de este modo generar la lista o catálogo de ficheros que inicialmente necesita el sistema. Mediante la utilidad *readahead* el SO alojado guarda un registro o catálogo de ciertos ficheros (memoria caché) que necesita en el proceso de arranque. Esto permite que cuando se inicie el SO, sólo accederá al citado registro de archivos y por tanto obtener como consecuencia una reducción del tiempo de carga del mismo SO.

El interés que se muestra en este trabajo por el paquete *readahead* se centra en la generación de información en el arranque del sistema. Esto implica conocer qué ficheros se necesitan en primera instancia para que el SO alojado de una imagen virtual pueda iniciar los recursos del sistema.

**3.2 Monitorizar la ejecución de una aplicación.**

En un SO alojado en una imagen virtual se pueden tener instaladas todas las aplicaciones informáticas que sean necesarias para cubrir el objetivo o funcionamiento de la propia MV. Estas aplicaciones o proyectos no siempre usan completamente todos los recursos del SO, al contrario, dependiendo de sus características o funcionalidad usarán ciertas partes o utilidades del sistema, unas más que otras. Es posible que incluso algunas aplicaciones no usen nunca algunas partes del sistema.

El segundo punto de ejecución objeto de estudio corresponde a la monitorización de una sesión habitual o más concretamente en el uso de algún aplicativo implementado en el SO alojado: concretamente haciendo uso de la herramienta software *preload*[7] (su funcionamiento es análogo al paquete *readahead*). El objetivo de esta monitorización permite obtener información sobre los recursos del sistema necesarios para la ejecución de una tarea o aplicación en un SO. Para explicar esta monitorización podría escogerse cualquier aplicativo ejecutable por el SO a nivel de usuario. En el presente trabajo se ha elegido un conocido paquete ofimático como es *OpenOffice*[8]. La idea es entender esta monitorización como si de un proyecto científico pudiera tratarse y que estuviera integrado, por ejemplo, en una infraestructura de tipo Desktop Grid (ambiente distribuido), formando parte de una unidad de trabajo en computación paralela o Computación Voluntaria. Durante un tiempo de ejecución determinado de *OpenOffice*, con la utilidad *preload* (previamente instalada y ejecutada en segundo plano), se va a generar una lista de archivos utilizados en esa sesión.

Con carácter general la ejecución de una aplicación instalada en un SO necesita todo el software correspondiente a la misma: librerías, imágenes, archivos de configuración, etc. Pero es posible que un SO no necesite todos sus recursos o servicios para ejecutar sus diversas tareas en el sistema, incluidas las relacionadas con aplicaciones de usuario concretas que pudiera tener instaladas. Esto último es principalmente lo que se pretende analizar, es decir, qué parte del SO son más necesarias y cuáles no se usan habitualmente por parte del usuario.

**3.3 Cálculo del espacio ocupado por el SO alojado.**

Antes de realizar las monitorizaciones comentadas anteriormente, es preciso calcular el espacio ocupado por el SO, es decir, de su sistema de ficheros[9]. La tabla 1 presenta los resultados obtenidos sobre una imagen

---

6   Fedora Hosted. Start a project.   https://fedorahosted.org/readahead/
7   SourceForge: preload.   http://sourceforge.net/projects/preload/
8   OpenOffice. http://www.openoffice.org/

virtual de 6 Gbytes[10] cuando ha sido reducida al tamaño exclusivo del SO alojado (incluyendo las aplicaciones que pudiera tener implementadas):

| SO alojado (*GNU/Linux*) | FS | SO anfitrión (*GNU/Linux*) | FS | Tamaño FS | % |
|---|---|---|---|---|---|
| Debian 5 (Lenny) | ext3 | Debian 5 (Lenny) | ext3 | 2,6 GB | 43,33 |
| Ubuntu 10.04 LTS (Lucid Lynx) | ext4 | Ubuntu 10.04 LTS (Lucid Lynx) | ext4 | 2,2 GB | 36,67 |
| OpenSuse 11.3 | ext4 | Ubuntu 10.04 LTS (Lucid Lynx) | ext4 | 3,3 GB | 55 |
| Fedora 10 (Cambridge) | ext3 | Debian 5 (Lenny) | ext3 | 3,5 GB | 58,33 |

Tabla 1: Características de los SO alojados y tamaño de los FS correspondientes.

Sobre la columna "**Tamaño FS**", se realizarán los cálculos posteriores de *readahead* y *preload*, es decir, la ocupación o tamaño que representa sobre el SO alojado (sistema de ficheros). La relación media de estos cuatro SO alojados sobre el total de espacio físico asignado en cada MV (6 GBytes) es del **48,33%** (**2,9 GBytes**), es decir, que más de la mitad del espacio asignado a la MV no es usado inicialmente por el SO alojado.[11]

**3.4 Resultados obtenidos en el arranque:** *readahead.*

La tabla 2 refleja la cantidad de archivos y ocupación que necesita un SO alojado, para realizar el arranque de una sesión habitual de un usuario del sistema (paquete *readahead*). Las monitorizaciones realizadas corresponden a un SO alojado con la distribución *GNU/Linux Debian Lenny*[12]:

| SO alojado (*GNU/Linux*) | Nº Ficheros accedidos | Tamaño (KB) | % sobre FS |
|---|---|---|---|
| Debian 5 (Lenny) | 578 | 36.817 | 1,35% |
| Ubuntu 10.04 LTS (Lucid Lynx) | 95<br>357 | 9.161<br>24.773 | 1,47% |
| Fedora 10 (Cambridge) | 1.315 | 158.441 | 4,32% |

Tabla 2: Cantidad (KBytes) del SO alojado requeridos por *readahead*

---

9   Para ello se hace uso de un procedimiento de montaje de particiones. El sistema de ficheros de un SO alojado en una MV puede ser migrado a un directorio del SO anfitrión y posteriormente realizar las tareas de observación o de análisis que se pretendan sobre ese nuevo directorio de montaje.
10  Se escoge este tamaño estimando el tamaño medio normal de un SO, así como el espacio en disco que habitualmente necesita para gestionar sus aplicaciones y recursos un SO. No obstante este dato no es relevante en el objetivo principal de este trabajo ya que como se indica en esta misma sección la reducción del SO alojado se realiza sobre el tamaño real del sistema de ficheros.
11  Evidentemente si el tamaño predeterminado para la imagen virtual original fuera distinto de 6 GBytes, la reducción inicial de la imagen también presentaría porcentajes variados ya que se trata en principio de aislar la imagen virtual (como archivo) del sistema de ficheros correspondiente al SO alojado.
12  Dependiendo de la distribución *GNU/Linux* instalada en la MV el SO necesitará mayor o menor número de archivos para realizar sus tareas, aunque esto depende también de la versión del software *readahead* implementada en el propio SO.

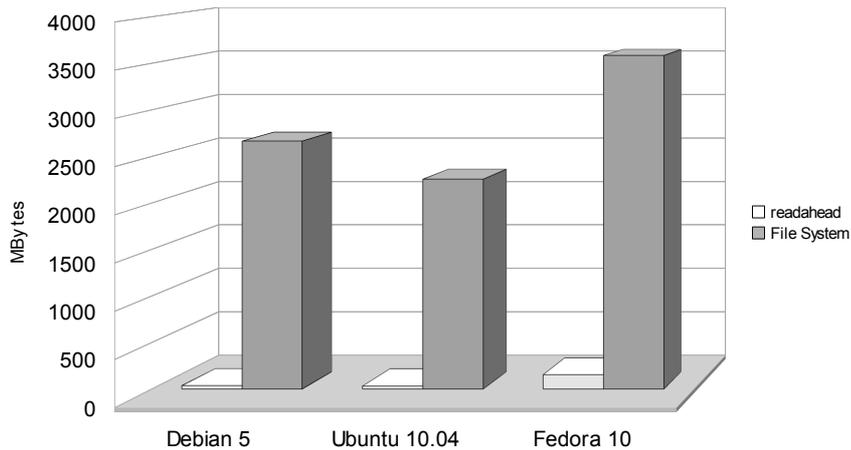

Figura 1: Comparativa entre el espacio ocupado por el FS y el arranque del SO alojado

La columna "**% sobre FS**" representa el tamaño del catálogo registrado por *readahead* sobre el tamaño real del sistema de ficheros (FS) que figuran en la columna "**Tamaño FS**" de la tabla 1. Los porcentajes que se muestran en dicha columna indican que muy poca información es usada por el SO en el momento del arranque, concretamente supone de media el **2,38%** sobre las distribuciones *GNU/Linux* instaladas en las MV analizadas.

**3.5 Resultados obtenidos en la ejecución de una aplicación:** *preload.*

Las monitorizaciones realizadas corresponden a un SO alojado con la distribución *GNU/Linux Ubuntu 10.04 LTS Lucid Lynx*. El aplicativo propuesto es el comentado anteriormente: *OpenOffice*. Los resultados obtenidos están representados en la tabla 3 y son los siguientes:[13]

| SO alojado (*GNU/Linux*) | Nº Ficheros accedidos | Tamaño (KB) | % sobre FS |
|---|---|---|---|
| Debian 5 (Lenny) | 781 | 278.597 | 10,22% |
| Ubuntu 10.04 LTS (Lucid Lynx) | 1.674 | 519.609 | 22,52% |
| OpenSuse 11.3 | 2.518 | 258.696 | 7,48% |
| Fedora 10 (Cambridge) | 1.127 | 435.905 | 11,88% |

Tabla 3: Cantidad (KBytes) del SO alojado requeridos por *preload*

De manera análoga a *readahead* y en relación a la tabla 1, los porcentajes muestran que la información usada por el SO en una sesión con un aplicativo ejecutándose, concretamente supone de media el **13%** sobre las distribuciones *GNU/Linux* instaladas en las MV.

---

13 El número de ficheros accedidos y el tamaño total depende de cada distribución *GNU/Linux* implementada, es decir, de la instalación inicial en cada caso. Esto explica las diferencias que existen en cada una de las monitorizaciones realizadas. El objetivo de esta sección es estimar lo más aproximado posible el tamaño de ficheros necesario requerido por el SO alojado en estas monitorizaciones y para ello se han elegido diversos SO en los que realizarlas.

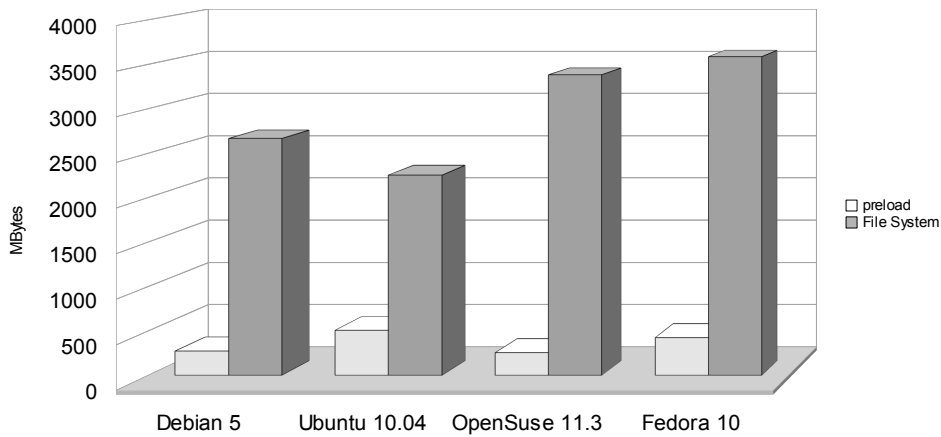
Figura 2: Comparativa del espacio usado por el FS y una sesión de usuario

Una buena aproximación al tamaño del catálogo de archivos referidos al *preload* consistiría en disponer exactamente de la aplicación científica que se pretenda evaluar y ejecutar todas las funcionalidades posibles incluidas en ella. Los datos sobre el número de ficheros accedidos y el tamaño final en bytes, sería más realista y estaría seguramente un poco más incrementado que el que se presenta en la tabla 3. No obstante cabe destacar que una misma aplicación científica puede usar algunos archivos de control y gestión de manera compartida en el SO, por lo que es posible que en estas monitorizaciones estarían implicados archivos comunes del sistema y particularmente del propio aplicativo científico que se monitorice.

## 3. Extracción de los datos monitorizados en el SO alojado.

Una vez analizado el funcionamiento del SO alojado en dos puntos de ejecución diferentes, el paso final consistiría en disponer de esa información monitorizada. Para llevar a cabo la extracción de datos del SO alojado en una MV, se han de realizar los siguientes pasos:

- **Localización del sistema de ficheros** del SO alojado en una máquina virtual.
- Permitir al SO anfitrión **disponer de los datos monitorizados** en el sistema de ficheros correspondiente al SO alojado.
- **Extraer la información monitorizada** y con la posibilidad de empaquetarla en un formato adecuado para su transferencia o distribución.

**4.1 Localización del sistema de ficheros del SO alojado.**

La herramienta de virtualización utilizada en el presente trabajo es *VirtualBox*[14]. En esta MV un archivo con formato VDI (Virtual Disk Image) representa físicamente una imagen virtual (extensión *.vdi*). Localizar el sistema de ficheros de un SO alojado implica conocer en qué parte de la imagen virtual se encuentra la zona de datos y por tanto conocer la geometría de una imagen virtual (estructura lógica y física) lo cual permitirá posteriormente extraer información del SO alojado en la misma.
La figura 3 muestra la estructura de una imagen virtual para la máquina VirtualBox[15]:

---

14  http://www.virtualbox.es/ Se trata de un software de virtualización para arquitecturas x86 por medio de la cual es posible instalar un SO invitado, dentro de otro SO anfitrión (host anfitrión). En este trabajo se han instalado MV *VirtualBox* (SO alojado) con varias distribuciones de *GNU/Linux* así como para cada SO anfitrión. Las imágenes virtuales (archivos físicos) correspondientes a máquinas de tipo *VirtualBox* tienen la extensión *.vdi*. Existen actualmente varias herramientas software que gestionan imágenes *.vdi* en *VirtualBox*. Gracias a estas utilidades y a la flexibilidad de *VirtualBox*, ha sido posible el análisis de las imágenes virtuales.
15  VX Heavens: VirtualBox's Virtual Disks Infection. http://vxheavens.com/lib/vwg03.html

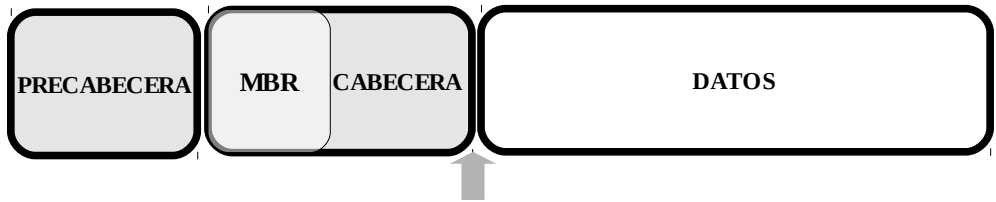

Figura 3: Estructura de una imagen *VirtualBox*.

El hipervisor (host anfitrión) hace uso de la información existente en la precabecera y la cabecera (datos de control) para gestionar la imagen (memoria, espacio de almacenamiento, etc.). En estos bloques de información se encuentran definidas las características de funcionamiento de la MV. La flecha color gris muestra el inicio de la zona de datos del SO alojado. En esta parte de la imagen virtual se encuentra el sistema de ficheros dónde se incluyen todos los archivos del SO. Por tanto es necesario establecer un proceso que permita extraer esta zona de datos, ya que es dónde se encontrarán incluidos los ficheros registrados o monitorizados por los procesos *readahead* y *preload*.

**4.2 Migración del SO alojado al SO anfitrión.**

Una vez localizado el sistema de ficheros del SO alojado ahora el segundo paso consiste en obtenerlo en un formato legible al SO anfitrión. Disponer de esta información en el host anfitrión permite la posibilidad de gestionarla y poder realizar con ella las operaciones que se necesiten. Concretamente sería necesario realizar los siguientes pasos desde el host anfitrión:

1) Convertir la imagen virtual original a un formato legible por un sistema de ficheros estándar, por ejemplo en el formato conocido como *raw*.
2) Separar la precabecera y cabecera de la imagen virtual de la zona de datos del SO alojado. Esto permite aislar los datos del SO alojado del resto de los datos de control correspondientes a la imagen virtual original. El resultado de la aplicación de estos dos pasos sobre la imagen virtual original es una nueva imagen resultante en formato *raw*. Esta nueva imagen *raw* contiene únicamente el sistema de ficheros correspondiente al SO alojado en la imagen virtual.
3) Finalmente se procede al montaje o migración de la imagen (generada en el paso anterior) hacia el host anfitrión.

La figura 4 representa gráficamente lo que se pretender realizar con el sistema de ficheros del SO alojado. Sus datos son extraídos hacia un directorio del host anfitrión. Posiblemente existan otros procedimientos para migrar información desde un SO alojado hacia su SO anfitrión (Monitor de Máquina Virtual, MMV), sin embargo la ventaja de realizar las operaciones mostradas en esta sección desde el host anfitrión, es la posibilidad de diseñar un procedimiento automático que permita migrar todas las imágenes virtuales de todas las MV alojadas en dicho host anfitrión hacia otros donde se desee transferir o distribuir.[16]

---

16  Esta opción de montar directorios del SO alojado hacia el SO anfitrión es también muy usada para realizar operaciones de recuperación de archivos sobre imágenes virtuales.

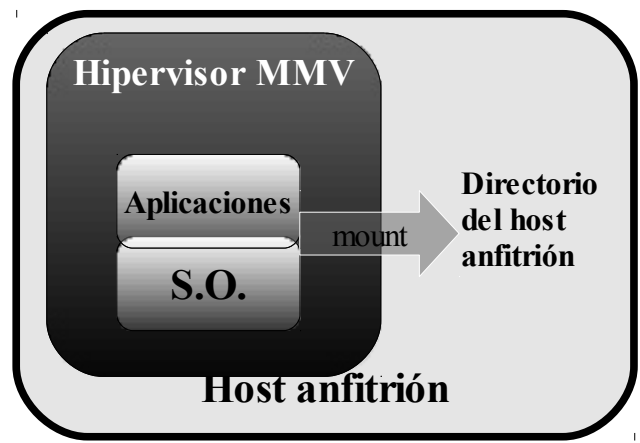

Figura 4: Sistema de ficheros (datos) de una MV: montaje en un directorio del host anfitrión

El procedimiento realizado en esta sección es indispensable para extraer la información monitorizada en el SO alojado. Esta información es la resultante de ejecutar las aplicaciones o servicios *readahead* y *preload* en la MV y con ella se pretende estudiar la posibilidad de reducir dicha imagen virtual con un SO de tamaño mínimo y al mismo tiempo personalizando el funcionamiento de la propia MV.

**4.3 Extracción de datos y generación de instancias.**

Finalmente en esta sección se comentará el procedimiento consistente en la extracción de datos del SO alojado. Constituye el paso final para obtener la reducción de la imagen virtual original.

En el paso presentado en la sección anterior en el host anfitrión ya se dispone de un directorio donde se encuentra el sistema de ficheros completo del SO alojado. Ahora para extraer la información monitorizada se deben realizar los siguientes pasos:

- Copiar en el host anfitrión los catálogos o registros de ficheros resultantes de la monitorización de *readahead* y *preload.*
- Finalmente extraer y empaquetar desde el directorio de montaje del host anfitrión los ficheros incluidos en cada uno de los catálogos referenciados en el paso anterior y que forman parte del sistema de ficheros del SO alojado. Por tanto, de este proceso se obtienen dos fragmentos diferentes (**instancias**) de lo que sería la imagen final reducida sobre la imagen original.

El resultado final en la realización de estos dos pasos es la obtención de dos archivos o fragmentos que incluyen los diferentes archivos, directorios, etc. que el SO alojado ha necesitado para la ejecución de los procesos *readahead* y *preload*.

Por tanto el resultado final de esta extracción equivale a dos partes del SO alojado que concretamente coinciden con el contenido de los catálogos obtenidos en los puntos de ejecución monitorizados en dicho SO, es decir, las utilidades o servicios *readahead* y *preload*.

# 1. Conclusiones.

En el presente trabajo se ha analizado una propuesta para la mejora en la transferencia o distribución de imágenes virtuales. Concretamente se basa en la monitorización del SO alojado en la MV y determinar qué partes del mismo están directamente implicadas con un funcionamiento específico para dicha máquina virtual.

Confeccionar una imagen virtual reducida con estas partes del SO alojado previamente monitorizadas representa de manera implícita una mejora en la transferencia de las imágenes virtuales (latencia, ancho de banda, etc.).

El resultado de estos tres pasos: "**monitorizar-extraer-instanciar**" resultaría en un nuevo SO (que podría alojarse en una nueva máquina virtual) donde se incluyeran los fragmentos obtenidos del SO alojado original y tal y como ha podido reflejarse en las tablas 2 y 3 suponen una diferencia de tamaño muy considerable con respecto a la imagen virtual inicial.

Las conclusiones del trabajo actual se basan en los resultados finales obtenidos una vez monitorizadas las actividades del SO alojado. Cada una de estas monitorizaciones ha registrado un porcentaje de utilización del sistema de ficheros, lo que se puede traducir igualmente a una ocupación física de archivos del SO, en la imagen virtual.

Gráficamente el mecanismo presentando se podría resumir en la figura 5:

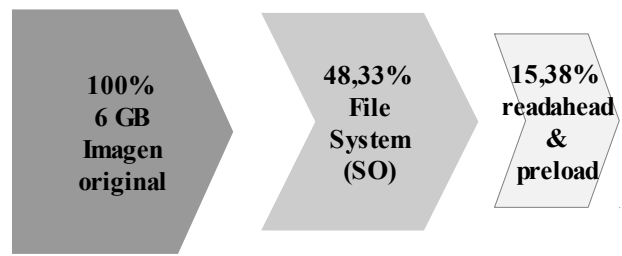

Figura 5: Resultados de la monitorización *readahead* y *preload*

Según los resultados mostrados en las tablas 2 y 3 se estaría concluyendo que el SO alojado usa el **15,38%** de su sistema de ficheros (no del tamaño original de la imagen virtual). Tal y como se refleja en los datos presentados en las tablas 2 y 3 los archivos necesarios en cada caso podrían reducir enormemente el tamaño de la imagen virtual final[17]. No obstante se han de considerar los recursos del SO alojado que no hayan podido incluirse en estos archivos de registro y que estén directamente relacionados con el funcionamiento normal donde se va a ejecutar la MV final reducida. Las utilidades *readahead* y *preload* recogen los recursos más usados por parte del sistema pero seguramente haría falta añadir algunas comprobaciones para concluir que esos recursos son los que se van a usar de manera exclusiva.

**Monitorización SO alojado = *readahead* (2,38%) + *preload* (13%)**

Partiendo de los datos presentados en la tabla 1 el sistema de ficheros de un SO alojado tiene un tamaño medio de 2,9 GBytes. Aplicar la fórmula anterior a este tamaño supone que el **15,38%** de esta cantidad es **457 Mbytes**.

Con las instancias obtenidas (fragmentos obtenidos por las monitorizaciones realizadas por *readahead* y *preload*) y que representarían aproximadamente este tamaño, se podría pensar en la posibilidad de diseñar un mecanismo de generación de imágenes virtuales (a partir de la información obtenida del SO alojado). El resultado podría ser imágenes con formato ISO Bootable[18] incluyendo toda la información necesaria (SO y aplicaciones asociadas) a modo de *DVD/CD/USB Live* o bien generar una imagen ISO Bootable semejante al punto anterior pero con la posibilidad de ligarlo a una MV que ejecutaría el SO alojado sobre dicha imagen ISO. Como consecuencia de la aplicación de este mecanismo (instancias o fragmentos de reducido tamaño) tendría un costo menor de transferencia que el de la imagen virtual original y al mismo tiempo debería disponer de la misma funcionalidad de la MV que se ha pretendido transferir en el entorno distribuido.

---

17 Los datos obtenidos pretenden ser una aproximación al tamaño final que tendría la imagen correspondiente a la MV. No es objeto de este trabajo estudiar a fondo, qué otros recursos serían necesarios por parte del SO alojado y que las utilidades *readahead* y *preload* no hubieran podido registrar.
18 Una ISO que incluya un SO con posibilidad de arrancar el propio sistema.